# Learning Assistant Supported Student Outcomes (LASSO) study initial findings

Ben Van Dusen[1], Laurie Langdon[2], and Valerie Otero[2]

[1]California State University Chico, Science Education Department.
Holt 101, Chico, CA, 95929, USA

[2]University of Colorado Boulder, School of Education
249 UCB, Boulder, CO, 80309, USA

**Abstract.** This study investigates how faculty, student, and course features are linked to student outcomes in Learning Assistant (LA) supported courses. Over 4,500 students and 17 instructors from 13 LA Alliance member institutions participated in the study. Each participating student completed an online concept inventory at the start (pre) and end (post) of their term. The physics concept inventories included Force and Motion Concept Evaluation (FMCE) and the Brief Electricity and Magnetism Assessment (BEMA). Concepts inventories from the fields of biology and chemistry were also included. Our analyses utilize hierarchical linear models that nest student level data (e.g. pre/post scores and gender) within course level data (e.g. discipline and course enrollment) to build models that examine student outcomes across institutions and disciplines. We report findings on the connections between students' outcomes and their gender, race, and time spent working with LAs as well as instructors' experiences with LAs.

## I. INTRODUCTION

The Learning Assistant (LA) model was developed for several reasons, including to improve undergraduate STEM student learning outcomes by increasing faculty use of research-based instructional strategies in undergraduate courses [1]. Since the introduction of the first LA workshop in 2007, the number of institutions with LA programs has grown from 3 to over 70 institutions [2]. In response to this growth, a coalition of LA using institutions (LA Alliance) was created. With the rapid growth of LA programs, there is a need for a list of LA program best-practices that new (and established) LA programs can draw from to ensure that their LA programs are effectively improving student outcomes.

Each of the 70 institutions in the LA Alliance has their own contextual affordances and constraints that act to shape the ways they implement their LA model. Even within a given institution, variation in classroom contexts, such as whether they have labs, what discipline they are teaching, and their LA/student ratio, can lead instructors to use LAs in significantly different ways. The uniqueness of each institutional and classroom context makes it difficult to identify specific aspects of LA programs that improve student outcomes and are replicable across institutions. The creation of the LA Alliance, however, made it possible to collect data across institutional settings. By examining LA-supported student outcomes across institutional and classroom contexts the LA Alliance is generating an evidentiary basis to create a list of context-specific best-practices for the use of LAs. In this paper we examine broad trends in student outcomes from the first semester of a multi-year study. The effects of LAs on student learning that are identified in this paper will act as a baseline for comparing the effects of specific LA-practices in future studies.

## II. RESEARCH QUESTIONS

By examining student outcomes, student demographics, and classroom features we investigate the questions: (1) How do teachers' uses of LAs predict student performance in LA supported courses, if at all? (2) How do course attributes predict student performance in LA supported courses, if at all? (3) How do students' attributes predict student performance in LA supported courses, if at all? (4) How do students' interactions with LAs predict student performance in LA supported courses, if at all?

## III. LITERATURE REVIEW

Researchers have documented the effects of LAs on students' conceptual learning in a number of institutional and classroom contexts. For example, one investigation examined the pre and post scores of approximately 5,000 physics students on physics concept inventories over a several year period [3]. It was found that student outcomes in introductory physics courses were significantly improved through the introduction of LAs and Tutorials [4]. Similar outcomes were found in a study of the effects of LAs with an alternative set of Tutorials on students in another institution [5]. In addition to improving learning gains, the use of LAs with tutorials was also shown to not exacerbate the performance gap between underrepresented student groups and majority groups. In a study of calculus students' course grades, the use of LAs was shown to completely close the performance gap between students' who had been labeled "at risk" (due to low entry exam scores) and the rest of the class [6]. Similar trends have been found in undergraduate chemistry classes. For example, in a multi-year study of an introductory chemistry class, it was found that introducing LAs while holding the curriculum constant led to significant differences in students' learning outcomes [7].

These kinds of findings have been a driver of the adoption of LA programs internationally, yet they provide no generalizable evidence about how to best use LAs. As these studies took place in single institutions that were typically introducing several changes (e.g. introducing LAs and tutorials) simultaneously there is currently way to distinguish the effects of particular interventions or know how they will vary across settings. In this study we look beyond any individual instantiation of the LA program to examine student outcomes across classroom, discipline, and institutional contexts. Using this broad set of data, we investigate how LAs affect student outcomes across learning environments.

## IV. METHODS

### A. Data Sources and Collection

Faculty were recruited for the study through two primary methods: (1) participants in the 2014 International LA Workshop were recruited through a series of research sessions and (2) The LA program coordinator at each site in the LA Alliance was sent an invitation email inviting their faculty to participate. To participate each faculty member was required to complete a brief online questionnaire about the features of their course (e.g. what discipline it covers, how many students are enrolled, are there mandatory recitations, etc.) and how they use LAs (e.g. in lectures, in recitations, etc.). The faculty were provided an email with directions for their students and a link to one of nine surveys that we hosted online through Qualtrics. Each survey begins with questions about the student (e.g. student ID, race, and how much time they spent interacting with LAs) and ended with the concept inventory the faculty had selected. The concept inventories faculty chose from included: Concept Inventory of Natural Selection (CINS) [8], Genetics Concept Assessment (GCA) [9], Introductory Molecular and Cell Biology Assessment (IMCA) [10], Chemistry Concept Inventory (CCI) [11]; General Chemistry II Concept Survey (GCIICS) [12], Force and Motion Concept Evaluation (FMCE) [13], Brief Electricity and Magnetism Assessment (BEMA) [14]. Students completed the surveys at both the start (pre) and end (post) of their terms.

Data was also collected from three classes that administered the concept inventories in class. There is no demographic information for the 1,570 (47% of the total sample size) students who completed the concept inventories in class, so they were excluded from any analyses that utilized student demographic variables.

In exchange for participating each faculty member was provided a report that showed the distribution of their students' pre and post scores, normalized learning gains (Hake score), and effect size (Cohen's d). Faculty were also provided an anonymized report that outlined the learning outcomes of students from all participating institution as well as how LAs were used in those courses.

In the pilot semester of data collection, a total of 8,654 unique student concept inventory score were collected from 17 courses in 13 institutions.

### B. Data Analysis

The first step in analyzing our pilot data was the cleaning of student responses. Student survey responses were cleaned in a two-ways. First, any survey with answers to less than 80% of the concept inventory questions was removed. Second, any student responses that were not part of a matching pre-post survey were removed. Once student results were cleaned, there were 3,315 usable pre-post pairs of responses. Each of these responses were scored and an effect size (Cohen's d) was calculated for each student. Cohen's d is a measure of change (in this case from pre to post scores) in units of standard deviations.

Student and faculty responses were used to generate Hierarchical Linear Models (HLM). HLM is typically used as a method of correcting for the obvious dependences created in nested data situations [23]. For example, honors physics courses are typically taught differently than algebra-based physics courses for non-majors. Differences in the ways that these courses are taught will likely have effects on students' performances on concept inventories. This dependence violates the assumptions of normal Ordinary Least Squares regression that all units are independently affected by outside stimuli. HLM can account for these differences by allowing for classroom level dependencies. In effect, HLM creates unique equations for each classroom and then uses those classroom-level equations to model an effect estimate across all classrooms.

In our HLM models we nested student level data within course level data. This allowed us to examine the correlations between student outcomes (i.e. effect sizes) and various factors across concept inventories. All of the models used effect size as the outcome, student data in the first level, and course data in the second level. These models took the generalized form:

Lvl 1 (student): $\text{Eff.Size}_{ij} = \beta_{0j} + \beta_{1j}(\text{Stud.Var.1}_{ij}) + \ldots + r_{ij}$
Lvl 2 (course): $B_{0j} = \gamma_{00} + \gamma_{01}(\text{CourseVar.1}_j) + \ldots + u_{0j}$
$\qquad\qquad B_{1j} = \gamma_{10} + \gamma_{11}(\text{CourseVar.1}_j) + \ldots + u_{1j}$
$\qquad\qquad \ldots$

In our analysis we created 8 models that examined different sets of variables: (1) No variables (an unconditional model), (2) *course level* – concept inventory variables (note: concept inventory variables were included as fixed effects in all of the other models that use course level data), (3) *student level* - gender variables, (4) *student level* - race variables, (5) *student level* – student interaction time with LAs (min/week), (6) *course level* - student to LA ratios, (7) *course level* – faculty meeting time with LAs (min/week), and (8) *course level* - number of times an instructor has taught the course with LAs.

## V. FINDINGS

The outcomes for the unconditional model (1) and the models that use categorical data (2-5) are shown in table 1. In models 3-5 tests of statistically significant differences were measured (p-values). In model 3 (gender) each variable was compared against the "male" variable. In model 4 (race) each variable was compared against "white" variable. In model 5 (student weekly interaction time with LAs) each variable was compared against "0 min" variable.

**TABLE 1.** HLM models with categorical variables.

| | Variable (N) | Average effect size | Standard error |
|---|---|---|---|
| **Model 1: Unconditional** | Grand mean (3,315) | 0.845 | 0.140 |
| **Model 2: Concept Inventory** | IMCA (2) | 0.306 | 0.284 |
| | GCIICS (1) | 0.591 | 0.375 |
| | GCA (1) | 0.207 | 0.447 |
| | FMCE (5) | 1.230 | 0.173 |
| | CINS (1) | 1.248 | 0.426 |
| | CCI (2) | 0.052 | 0.268 |
| | BEMA (5) | 1.130 | 0.191 |
| **Model 3: Gender** | Female (944)** | 0.485 | 0.046 |
| | Male (789) | 0.749 | 0.050 |
| | Transgender (8) | 0.417 | 0.495 |
| | Other (3) | 0.349 | 0.809 |
| **Model 4: Race** | White (1,235) | 0.619 | 0.040 |
| | Asian (282)** | 0.557 | 0.084 |
| | Black (77)$^\tau$ | 0.669 | 0.160 |
| | American Indian (7) | 0.849 | 0.531 |
| | Haw./Pac. Isl. (7) | 0.706 | 0.531 |
| | Other (127)** | 0.509 | 0.125 |
| **Model 5: Student weekly interaction time with LAs** | 0 min (280) | 0.369 | 0.085 |
| | 1-5 min (299) | 0.517 | 0.082 |
| | 6-15 min (318) | 0.663 | 0.080 |
| | 16-30 min (184)** | 0.873 | 0.105 |
| | 30+ min (203) | 0.569 | 0.100 |

$\tau$ p<0.1, * p<0.05, ** p<0.005

*Model 1* shows that the grand mean for student effect sizes is 0.845. This means that on average students scored 0.845 standard deviations higher on their post-tests than their pre-tests. *Model 2* shows that average student effect sizes by concept inventory ranged from 1.248 (CINS) to 0.052 (CCI). The average effect size for each concept inventory was used as a fixed effect to account for differences in course level data. *Model 3* shows that the average effect sizes for males (0.748) and females (0.485) students were statistically significantly different. *Model 4* shows that the average effect sizes for white students (0.619) were statistically significantly different from Asian (0.557), black (0.669), and "other" (0.509) students. In comparison to white students the model shows that the average effect size for black students was higher, while those of Asian and "other" students was lower. *Model 5* shows that students' average effect size increases with the amount of time they spend with LAs. The trend peaks at 16-30 minutes/week, which is also the only category to be statistically significant from the 0 minutes/week category.

The outcomes for models with continuous variables (6-8) are shown in Table 2. *Model 6* shows that as the student to LA ratio increases, there is a decrease in students' average effect size (-0.005 per additional student/LA). This means that students in classes with less students and more LAs do better. *Model 7* shows that student performance was better in classes where faculty spent more time meeting with LAs (+0.004 for every additional minute faculty spent co-planning with LAs). *Model 8* shows that student performance increased with faculty experience teaching the course with LAs. On average, for every term a faculty member had taught the course with LAs, their students' effect sizes were 0.154 higher (a statistically significant difference).

**TABLE 2.** HLM models with continuous variables (n=17).

| | Effect size/unit change in variable | Standard error |
|---|---|---|
| **Model 6: Student to LA Ratio** | -0.005 | 0.003 |
| **Model 7: Faculty meeting with LAs (min/week)** | 0.004 | 0.004 |
| **Model 8: Times taught course with LAs** | 0.154* | 0.065 |

* p<0.05

## VI. DISCUSSION

When examining the average effect sizes by concept inventory, one may notice that the average effect sizes on the two physics concept inventories (FMCE & BEMA) are relatively large compared to other disciplines. This does not mean that LA using physics instructors are, on average, any more effective teachers than their Biology and Chemistry counterparts. Student scores on concept inventories are subject to a number of influences, including the alignment of the curriculum with the instrument and the quality of the instrument itself. Physics has developing and using concept inventories for longer than the other sciences and it may be that this has acted to drive up the average physics student's effect size.

Because our hierarchical linear models nest the students' data within the classroom data, it allows us to control for disciplinary differences in the generation of our models. In the remainder of this discussion, we will only address the statistically significant findings.

Our findings showed that there were a number of gendered and racial inequities that mirror some of the same inequities that are endemic to the sciences. Male students had average effect sizes that were more than 50% higher than the average effect sizes for females. The racial data shows that one group of traditionally underserved students (black students) had higher average effect sizes than their white peers, but that the white students outperformed those who reported being Asian or "other". It should be noted that these findings are correlational and causation should not be inferred. The effects of LAs on these inequalities are topics for further study.

It is encouraging that, on average, as student interaction times with LAs increases so do their effect sizes (up to a point). Students who spent 16-30 minutes/week interacting with LAs had average effect sizes that were more than *double* those of students who spent 0 min/week interacting with LAs. If it is the interactions with LAs that are causing these shifts, then a simple weekly 16-30 minute intervention is more than doubling student performance. Interestingly, students who spent 30+ minutes/week working directly with LAs saw no improvement over students who spent 0 minutes/week working with LAs. One potential explanation for this finding is that there is a selection effect for student students who choose to spend 30+ minutes/week with LAs. It may be that these high LA-using students are the students who are struggling the most in the course. So even though LAs are having a positive effect on the high-LA using students' the difficulty of the course for these students reduces their average effect size.

It also encouraging that, on average, the longer an instructor has used LAs in a course the larger their students' effect sizes are. For *each* time an instructor has taught a class using LAs, their students average 0.154 standard deviation higher on their post-test. It is unknown for how many terms this trend will continue for, but the faculty in this sample ranged from never having taught the class with LAs to having taught it 6 times before with LAs. This finding supports previous hypotheses that LA programs act as platforms for creating and sustaining both faculty and institutional transformations [3,5]. If LAs are acting to transform instructor's pedagogical practices, it is no surprise that their student outcomes would improve over time.

## VII. CONCLUSION & FUTURE WORK

This study produced mixed findings. Unfortunately, there are gendered and racial inequities across LA supported classrooms. The findings that were specifically about interactions with LAs were highly positive. Statistically significant improvements in student outcomes were seen when LAs had sustained interactions with either students or faculty.

The findings presented in this paper are preliminary in nature are designed to act as a launching point for future investigations. The data analyzed in this study comes from the first term of a multi-year study. We are in the process of collecting additional data and merging our database with the database of researchers performing similar research [15]. With this growth in sample size there will be an increase in statistical power such that more nuanced hierarchical linear models can be generated. With these models, we intend to empirically address a number of research questions, including how specific uses of LAs and classroom features affect student outcomes.